\documentclass[letter,traditabstract]{aa}

\usepackage{amssymb}
\usepackage{amsmath}
\usepackage{graphicx}
\usepackage{natbib}
\usepackage{txfonts}
\usepackage[colorlinks=true,citecolor=blue,linkcolor=blue]{hyperref}
\usepackage{microtype}
\usepackage{color}

\bibpunct{(}{)}{;}{a}{}{,} 

\begin{document}

\title{GRB 090618: a Candidate of a Neutron Star Gravitational Collapse to a Black Hole Induced by a Type Ib/c Supernova}
\titlerunning{On Induced Gravitational Collapse of a NS by a Ib/c SN}
\authorrunning{L. Izzo, J. A. Rueda and R. Ruffini}

\author{L.~Izzo, Jorge A.~Rueda, and R.~Ruffini}
\institute{Dipartimento di Fisica and ICRA, Sapienza Universit\`a di Roma, P.le Aldo Moro 5, I–-00185 
Rome, Italy\\ICRANet, P.zza della Repubblica 10, I--65122 Pescara, Italy}

\offprints{\email{luca.izzo@icra.it, jorge.rueda@icra.it, ruffini@icra.it}}

\date{}

\abstract{
A novel concept has been recently proposed for explaining the temporal coincidence of some Gamma Ray Bursts (GRBs) with an associated Supernova (SN) in terms of the gravitational collapse of a neutron star (NS)  to a Black Hole (BH), induced by a type Ib/c SN explosion. We apply these considerations to the exceptional case of GRB 090618, for which there is evidence of a SN $\sim 10$ days after the GRB occurrence. We calculate the accretion rate and total accreted mass onto a NS from a SN Ib/c originated from a companion evolved star. It is shown that the NS reaches in a few seconds the critical mass and undergoes gravitational collapse to a BH leading to the emission of a GRB. We find for the mass of the NS companion, $M_{\rm NS}$, and for the SN core progenitor, $M_{\rm core}$, the following mass ranges: $1.8\lesssim M_{NS}/M_\odot \lesssim 2.1$ and $3\leq M_{\rm core}/M_\odot \leq 8$. Finally, we discuss the complementarity of these considerations to alternative processes explaining long and short GRBs.}

\keywords{Type Ib/c Supernova -- Binary system -- Neutron Star Accretion -- Gravitational Collapse -- Gamma Ray Bursts}

\maketitle

The temporal coincidence of some Gamma Ray Bursts (GRBs) and their associated Supernovae (SNe) represent one of the outstanding and hottest topic in the Relativistic Physics and Astrophysics. We recall that the optical emission of a SN reaches its maximum about 15 days after the SN explosion \citep[see e.g.][]{1996snai.book.....A}. This implies that the onset of SNe and the occurrence of the GRB can be almost simultaneous \citep[see e.g.][for a recent review on GRB-SN systems]{2011IJMPD..20.1745D,2011arXiv1104.2274H}. For these reasons the concept of the induced gravitational collapse (IGC) was initially proposed \citep{2001ApJ...555L.117R} and further developed in \citep{2008mgm..conf..368R}. Recently we have theoretically developed the basic equations describing this concept \citep{2012ApJ...758L...7R}. In \citep{2012ApJ...758L...7R} the GRB-SN connection has been explained in terms of the final stage of the evolution of a close binary system composed of an evolved star (likely a C+O star in the case of SN Ic) with a neutron star (NS) companion. The explosion of the SN Ib/c in this system leads to an accretion process onto the NS companion. The NS reaches the critical mass value in a few seconds, undergoing gravitational collapse to a Black Hole (BH). The process of gravitational collapse to a BH leads to the emission of the GRB. We have evaluated the accretion rate onto the NS, and we have given the explicit expression of the accreted mass as a function of the nature of the components and the binary parameters. 

%

We turn now to an explicit quantitative description of the process of IGC and compare the results with the observational data of GRB-SN. We apply in this letter these considerations to the specific case of GRB 090618 and its associated SN. We recall that GRB 090618 is one of the nearest ($z = 0.54$) and at the same time more energetic GRBs ($E_{iso} = 3\times 10^{53}$ erg), ever observed. In fact, no beaming appear to be present from the data. These two important properties of GRB 090618 have produced an exceptional set of data in view of the very fortunate joint observations of the major satellites observing in X and $\gamma$ rays: Fermi, Swift-BAT and XRT, AGILE, Suzaku-WAM, Konus-WIND and Coronas PHOTON RT-2. This has allowed to study GRB 090618 in a very wide electromagnetic energy range, from 0.3 keV to 40 MeV. There have been as well observations from the Optical telescopes: Hubble Space Telescope, Shane Telescope at the Lick Observatory and Palomar 60-inch Telescope, which led to the clear identification in the afterglow light curve of GRB 090618 of a late $\sim 10$ days optical bump associated to the SN emission \citep{2011MNRAS.413..669C}. 

\begin{figure}
\centering
\includegraphics[scale=0.2]{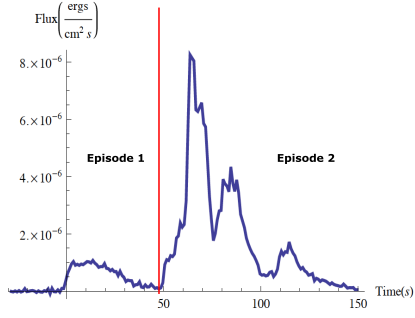}
\caption{Fermi GRB (NaI 8-1000 keV) light curve of GRB 090618. We explicitly indicate the two episodes of GRB 090618, identified by \cite{2012A&A...543A..10I}. The first episode lasting 50 s in the observed frame corresponds to $\sim 32$ s in the rest frame of this source.}\label{fig:lightcurve}
\end{figure}

\begin{figure}
\centering
\includegraphics[scale=0.34]{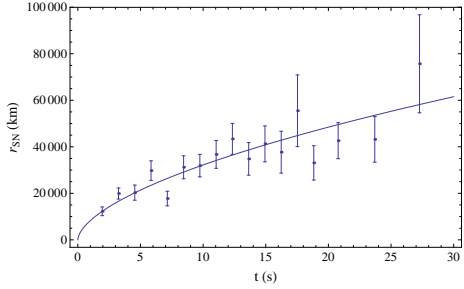}
\caption{Evolution of the radius of the black body emitter given by Eq.~(\ref{eq:rph}) in the first episode of GRB 090618 \citep{2012A&A...543A..10I}, here assumed as the radius of the SN. The time is here measured in the rest frame, so divided by a factor $1+z$, where $z=0.54$, with respect to the one of the first episode given in Fig.~\ref{fig:lightcurve}.}\label{fig:r_t}
\end{figure}

We have recently shown that GRB 090618 \citep{2012A&A...543A..10I} is composed by two sharply different emission episodes (see Fig.~\ref{fig:lightcurve}). From a time-resolved spectral analysis, it turns out that the first episode, which lasts $\sim 32$ s in the rest frame, is characterized by a black body emission that evolves due to a decreasing temperature with time (see Fig.~17 in \cite{2012A&A...543A..10I}). Associated to the decreasing black body temperature, the radius of the emitter has been found to increase with time (see Fig.~\ref{fig:r_t} and Fig.~18 in \cite{2012A&A...543A..10I}). With the knowledge of the evolution of the radius of the black body emitter, we find that it expands at non-relativistic velocities (see Eq.~(\ref{eq:rph}), below). Consequently, the first episode can not be associated to a GRB. Being prior to the GRB and therefore to the BH formation, this first episode emission has been temporally called Proto-BH, $\pi \rho \tilde{\omega} \tau o \varsigma$, from the ancient Greek that means anteriority in space and time. 

We here identify the Proto-BH of the first episode as the first stages of the SN expansion. The black body emitting surface in the first episode evolves during the first $\sim 32$ s, as observed in the rest frame, following a power-law behavior
\begin{equation}\label{eq:rph}
r_{\rm SN} = \sigma t^n\, ,\qquad v_{\rm SN} = n \frac{r_{\rm SN}}{t} = n \sigma t^{n-1}\, ,
\end{equation}
where $\sigma=8.048\times 10^8$ cm s$^{-n}$, $n\approx 3/5$ as shown in Fig.~\ref{fig:r_t}, and $v_{\rm SN}=dr_{\rm SN}/dt$ is the corresponding early SN velocity of the SN, so $\sim 4\times 10^8$ cm s$^{-1}$ at the beginning of the expansion.
%
%

When the accreted mass onto the NS triggers the gravitational collapse of the NS to a BH, the authentic GRB emission is observed in the subsequent episode at $t-t_0 \gtrsim 50$ s (observer frame). The characteristics of GRB 090618 are shown in Table 3 of \cite{2012A&A...543A..10I} and we refer to that reference for further details on the GRB light curve and spectrum simulation. We find that the optically thick $e^+e^-$ plasma expands reaching a bulk Lorentz factor $\Gamma \sim 500$ at a radius of $r_{tr}=1.4\times 10^{14}$ cm. At the transparency of the $e^+e^-$ plasma the proper GRB emission has been observed at a temperature of 29 keV and the corresponding baryon load, $B\sim 2\times 10^{-3}$, has been obtained (see Sec.~5.2 in \cite{2012A&A...543A..10I}). The baryon load is defined as $B=M_B c^2/E_{tot}^{e^+e^-}$, where $M_B$ is the engulfed baryon mass from the progenitor remnant and $E_{tot}^{e^+e^-}$ is the total energy of the $e^+e-$ plasma, which is assumed in the fit as the total observed GRB isotropic energy. We have obtained an average density of the circumburst medium (CBM) of 0.6 particles/cm$^3$, and each spike of the light curve corresponds to an interaction of the expanding plasma with clouds of radius $\approx 10^{15}$ cm and densities $\approx$ 2 particles/cm$^3$ (see Fig.~10 in \cite{2012A&A...543A..10I}, for details). 


We turn now to the details of the accretion process of the SN material onto the NS. The NS of initial mass $M_{NS}$ accretes mass from the SN ejecta at a rate given by \citep[see][for details]{2012ApJ...758L...7R}
\begin{eqnarray}\label{eq:deltaM}
\dot{M}_{acc}(t)= \pi \rho_{ej}(t) \frac{(2 G M_{NS})^2}{v^3_{\rm rel,ej}}\, ,\qquad \rho_{ej}(t)=\frac{3 M_{\rm ej}(t)}{4 \pi r^3_{\rm SN}(t)}\, ,
\end{eqnarray}
where $r^3_{\rm SN}(t)$ given by Eq.~(\ref{eq:rph}), $M_{\rm ej}(t)=M_{\rm ej,0}-M_{acc}(t)$ is the available mass to be accreted by the NS as a function of time, with $M_{\rm ej,0}$ the mass ejected in the SN. $v_{\rm rel,ej}=\sqrt{v^2_{orb}+v^2_{\rm SN}}$ is the velocity of the ejecta relative to the NS, where $v_{\rm SN}$ is the SN ejecta velocity given by Eq.~(\ref{eq:rph}) and $v_{\rm orb} = \sqrt{G (M_{\rm core}+M_{NS})/a}$ is the orbital velocity of the NS. Here $M_{\rm core}$ is the mass of the SN core progenitor and $a$ the binary separation. Hereafter we assume $a=9\times 10^9$ cm, a value larger than the maximum distance traveled by the SN material during the total time interval of Episode 1, $\Delta t\simeq 32$ s, $\Delta r \sim 7\times 10^9$ cm (see Fig.~\ref{fig:r_t}). 

In the case when the accreted mass onto the NS is much smaller than the initial mass of the ejecta, i.e. $M_{acc}/M_{\rm ej,0}<<1$, the total accreted mass can be obtained from the analytic expression given by Eq.~(8) of \citep{2012ApJ...758L...7R}, which in the case of GRB 090618 leads to
\begin{eqnarray}\label{eq:deltaMapp}
M_{acc}(t) &=& \left.\int_{t^{\rm acc}_0}^t \dot{M}_{acc}(t) dt\approx (2 G M_{\rm NS})^2\frac{15 M_{\rm ej,0} t^{2/5}}{8 n^3 \sigma^6 \sqrt{1+k t^{4/5}}}\right|_{t^{\rm acc}_0}^t\, ,
\end{eqnarray}
where $k=v^2_{orb}/(n\sigma)^2$ and $t^{\rm acc}_0$ is the time at which the accretion process starts, namely the time at which the SN ejecta reaches the NS capture region, $R_{cap}=2 G M_{NS}/v^2_{\rm rel,ej}$, so for $t \leq t^{\rm acc}_0$ we have $M_{acc} (t)=0$. The accretion process leads to the gravitational collapse of the NS to a BH when it reaches the critical mass value. Here we adopt the critical mass $M_{\rm crit}=2.67 M_\odot$ computed recently by \cite{belvedere2012}. The Eq.~(\ref{eq:deltaMapp}) is more accurate for massive NSs since the amount of mass needed to reach the critical mass by accretion is much smaller than $M_{\rm ej,0}$. In general, the total accreted mass must be computed from the numerical integration of Eq.~(\ref{eq:deltaM}), which we present below for the case of GRB 090618.

\begin{figure}
\centering
\includegraphics[scale=0.50]{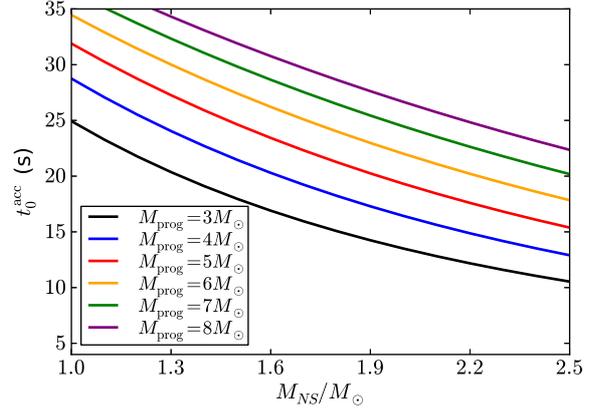}
\caption{Time $t^{\rm acc}_0$ since the SN explosion when the accretion process onto the NS starts as a function of the initial mass of the NS $M_{NS}$ and for selected values of the initial ejected mass $M_{\rm ej,0}$, in the case of GRB 090618.}\label{fig:t0acc}
\end{figure}



The occurrence of a GRB-SN event in the accretion induced collapse scenario is subjected to some specific conditions of the binary progenitor system such as a short binary separation and orbital period. The orbital period in the present case is
\begin{equation}\label{eq:period}
P=\sqrt{\frac{4\pi^2 a^3}{G (M_{\rm core}+M_{NS})}}=9.1 \left(\frac{M_{\rm core}+M_{NS}}{M_\odot}\right)^{-1/2}\,\,{\rm min}\, .
\end{equation}

\begin{figure}
\centering
\includegraphics[scale=0.50]{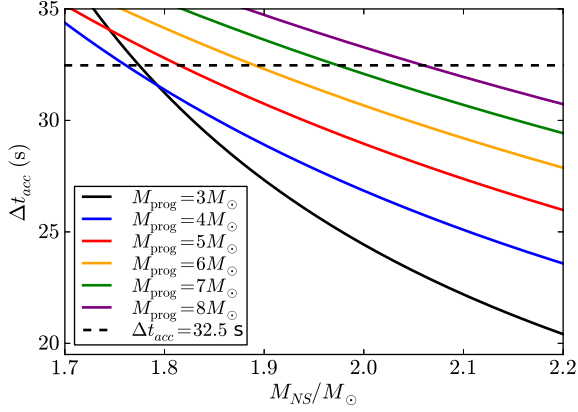}
\caption{Time interval $\Delta t_{acc}$ of the accretion process onto the NS as a function of initial NS mass $M_{NS}$ and for selected values of the SN core progenitor mass $M_{\rm core}$. The horizontal dashed line is the duration of the first episode of GRB 090618, $\Delta t=32.5$ s, which constraints the duration of the time needed by the NS to reach the critical mass. The crossing points between the dashed horizontal line and the solid curves give the NSs with $M_{NS}$ that reach the critical mass in the time $\Delta t$.}\label{fig:deltaT}
\end{figure}

We denote as $\Delta t_{\rm acc}$ the total time interval since the beginning of the SN ejecta expansion all the way up to the instant where the NS reaches the critical mass. In Fig.~\ref{fig:deltaT} we have plotted $\Delta t_{\rm acc}$ as a function of the initial NS mass, and for different masses of the SN core progenitor mass. The mass of the SN ejecta has been assumed to be $M_{\rm ej,0}=M_{\rm core}-M_{\rm rem}$, where $M_{\rm rem}$ is the mass of the central compact remnant (NS) left by the SN explosion. Here we have assumed $M_{\rm core}=(3$--$8) M_\odot$ at the epoch of the SN explosion, and $M_{\rm rem}=1.3 M_\odot$, following some of the type Ic SN progenitors studied in \citep{1988PhR...163...13N,nomoto1994,1994ApJ...437L.115I}.

We can see from Fig.~\ref{fig:deltaT} that, in the case of GRB 090618, the mass of the NS companion that collapses to a BH, should be in the range $1.8\lesssim M_{NS}/M_\odot \lesssim 2.1$ corresponding to the SN Ic progenitors $3\leq M_{\rm core}/M_\odot \leq 8$. 

The massive NS companion of the evolved star is in line with the binary scenario proposed by \cite{2008mgm..conf..368R}. It is appropriate to illustrate a more accurate evolutionary scenario that leads to the occurrence of a GRB-SN event following the works of \cite{1988PhR...163...13N,nomoto1994,1994ApJ...437L.115I}. The binary system is expected to evolve as follows: the initial binary system is composed of main-sequence stars 1 and 2 with a mass ratio $M_2/M_1 \gtrsim 0.4$. The initial mass of the star 1 is expected to be $M_1 \gtrsim 11 M_\odot$, evolving through a SN event leaving a NS. The star 2, now with $M_2 \gtrsim 11 M_\odot$ after some almost conservative mass transfer, evolves filling its Roche lobe. It then starts a spiral in of the NS into the envelope of the star 2. Consequently, it will be composed of a Helium star and a NS in close orbit. The Helium star expands filling its Roche lobe and a non-conservative mass transfer to the NS, takes place. This scenario naturally leads to a binary system composed of a C+O star and a massive NS, which can originate the process of IGC that we have proposed \citep{2008mgm..conf..368R,2012ApJ...758L...7R}.

These results are also in agreement with the well understood Ib/c nature of the SN associated with GRBs. The most likely explanation for SN Ib/c, which lack H and He in their spectra, is that the SN progenitor star is in a binary system with a NS,  see also \cite{1988PhR...163...13N,nomoto1994,1994ApJ...437L.115I} and also \cite{2007ARep...51..291T,2012ApJ...752L...2C}

In the current literature, the GRB-SN connection has been at time assumed to originate from a specially violent SN process called hypernova or a collapsar \citep[see e.g.][and references therein]{2006ARA&A..44..507W}. Both of these possibilities imply a very dense and strong wind-like CBM structure. This dense medium is in contrast with the CBM density found in most GRBs \citep[see e.g.][]{2011IJMPD..20.1797R}. Specifically, for GRB 090618 the average CBM density, $\langle n \rangle$, inferred from the analysis of the afterglow has been shown to be of $\sim 1$ particle/cm$^3$ (see e.g.~Fig.~10 and Table 3 in \cite{2012A&A...543A..10I}).The Lorentz Gamma factor of the electron-positron plasma has been estimated to be $\Gamma = 495 \pm 40$, corresponding to a baryon load $B = (1.98 \pm 0.15) \times 10^{-3}$, \cite[see Table 3 in][]{2012A&A...543A..10I}.  Such an ultra-relativistic expansion can only occur if the amount of engulfed baryon matter, quantifiable through the baryon load parameter $B$ does not exceed the critical value $B \sim 10^{-2}$ \citep{2000A&A...359..855R}. In view of the spherical symmetry of GRB 090618 this limit would be largely violated in the collapsar model. 

It is also interesting to compare the results on the IGC of a NS to a BH by a type Ib/c supernova \cite{2012ApJ...758L...7R} with the results of Chevalier \citep{1989ApJ...346..847C} on the accretion of a supernova material by the central NS generated by the supernova. A total accreted mass of up to $0.1 M_\odot$ in a time of a few hours was obtained there for normal type II SN. Thus, a similar amount of mass can be accreted in the two cases but in the latter the accretion occurs in a longer time. For reaching a high accretion rate of the inner SN material onto the central NS, there is the necessity of a mechanism that helps to increase the density of the NS surrounding layers, which is decreasing due to the expansion after being unbound by the SN explosion. \cite{1989ApJ...346..847C} analyzed the possibility of having a reverse shock wave that makes that work while it moves back through the SN core. The reverse shock is formed in the interaction of the mantle gas with the low density envelope. The time scale of the accretion process is thus determined by the time the reverse shock takes to reach the vicinity of the central newly born NS, which is of a few hours in the case of SN II progenitors. 
However, the existence of a low density outer envelope, e.g. H and He outer layers, is essential for the strength of the reverse shock. Fall back accretion onto the central NS is expected to be relevant only in SN II but not in SN Ic as the ones associated to GRBs, where H and He are absent. 
 
The argument presented in this letter naturally explain the sequence of events: SN explosion - IGC- BH formation - GRB emission.
Correspondingly, the accretion of the material ejected by the SN into the nearby NS of the IGC model presented here occurs almost instantaneously. In fact, for the SN expansion parameters obtained from the observations of Episode 1 in GRB 090618 (see Eq.~(\ref{eq:rph}) and Fig.~\ref{fig:r_t}), the accretion of the SN material onto the nearby NS occurs in a few seconds (see Figs.~\ref{fig:t0acc} and \ref{fig:deltaT}). The binary parameters are such that the ejecta density does not decrease too much (from $10^6$ to $\sim 10^4$ g cm$^{-3}$) before reaching the capture region of the NS, leading to a high accretion rate. As pointed out by \cite{1989ApJ...346..847C}, radiative diffusion will lower the accretion rate up to the Eddington limit (and then to even lower rates) when the trapping radius of the radiation in the flow $r_{tr}=\kappa \dot{M}_{acc}/(4\pi c)$ \citep[see Eq.~(3.24) in][]{1989ApJ...346..847C}, where $\kappa$ is the opacity, is equal to the Bondi radius $r_{B}=G M_{NS}/v^2_{\rm rel,ej}$, the gravitational capture radius. The radius $r_{tr}$ is located where the outward diffusion luminosity is equal to the inward convective luminosity. It can be checked that for the parameters of our system given by Eqs.~(\ref{eq:rph})--(\ref{eq:deltaMapp}), the equality $r_{tr}=r_B$ occurs in a characteristic time $\sim 200$ days, where we have used $\kappa = 0.2$ cm$^2$ g$^{-1}$. Thus, this regime is not reached in the present case since the NS is brought to its critical mass just in a few seconds. In the case analyzed by \cite{1989ApJ...346..847C}, it happens in a time $\sim 8$ days.

It is also appropriate, before concluding, to consider possible astrophysical systems which might have been the precursor of the GRB 090618 and its associated SN, as well as their possible final outcome.

As for the precursor, a natural candidate are binary X-ray sources formed by a NS in a binary system with a main sequence companion, where the masses of the NS are observed to be close to the canonical value $1.4 M_\odot$, e.g. Cen X-3 and Her X-1 \citep{1972ApJ...172L..79S,1972ApJ...174L..27W,1972ApJ...174L.143T,1973ApJ...180L..15L,1973ApJ...179..585D,1975ASSL...48.....G}; see also \cite{2011ApJ...730...25R}, for up-to-date estimates of the mass of the NSs in the X-ray binaries Vela X-1, 4U 1538-52, SMC X-1, LMC X-4, Cen X-3, and Her X-1.

As for the possible final outcome, let us turn to the final state after the explosion of the SN and the trigger of the GRB. There is still a possibility of forming a NS-BH binary system. The outcome of the SN explosion leads necessarily to the formation of a NS: evidence can be acquired by the observation of a characteristic late time X-ray emission (called URCA sources, see \cite{2005tmgm.meet..369R}). This emission has been interpreted as originated from the young ($t \sim$ 1 minute--$(10$--$100)$ years), hot ($T \sim 10^7$--$10^8$ K) NS, which we have called neo-NS \citep[see][for details]{2012A&A...540A..12N}. This emission can indeed be observed in the afterglow light curves of GRB 090618 \citep[see e.g.][]{2012A&A...543A..10I} and GRB 101023 \citep[see e.g.][]{2012A&A...538A..58P}. If the neo-NS and the BH are gravitationally bound after the GRB-SN emission, this new binary system might lead itself to a new merging process to a single BH. In this case the system could originate a further GRB with a possibly predominant emission in gravitational waves.
 
In conclusion, the IGC binary scenario applied here to the specific case of GRB 090618 naturally leads to the understanding of the energetics and the temporal coincidence of SN and GRBs, as well as their astrophysical scenario and makes new predictions on the final outcomes. It originates from a binary system composed of an evolved core and a NS. It is clear, however, that such GRB and their associated SN form a special class of long GRBs and of SNe Ib/c. There are in fact SNe Ib/c not associated to a GRB, e.g.~SN 1994I \citep{2002ApJ...573L..27I} and SN 2002ap \citep{2004A&A...413..107S}. Their observations refer to late phases of the SN evolution typically after $\sim 15$--$20$ days after the original collapse process. The existing descriptions of these late phases after 15-20 days from the original explosion make use of a Sedov type behavior $r\propto t^{2/5}$ \citep{sedov46,1959sdmm.book.....S}. In the present case of the IGC we are presenting here for the first time, the first $\sim 30$ s of the very early evolution of a SN Ib/c associated to a GRB (see Eq.~(\ref{eq:rph}) and Fig.~\ref{fig:r_t}). The energetic of this SN Ib/c, as shown from Episode 1, appears to be much higher than the ones of the usual SNe Ib/c not associated to GRBs, $E_{iso,Epi1} \propto 10^{52}$ erg \citep{2012A&A...543A..10I}. The reason of this marked difference is certainly due to the accretion process during a SN explosion into the companion NS and consequent gravitational collapse of the NS to a BH. The description of this challenging process although clear from a global energetic point of view, has still to be theoretically explored in details and certainly does not present any relation to the Sedov type  solution. 

From the point of view of the long GRBs, it is clear that the IGC concept leads to the formation of a BH of typical mass just larger than the critical mass of NS. 

There are of course GRBs originating in binary-NS systems and clearly do not have an associated SN, e.g.~GRB 090227B \citep{2012arXiv1205.6600M}. These binary systems clearly depart from the one studied in \citep{2012ApJ...758L...7R} and in this Letter. For selected masses of the NS, such binary system leads to a short GRB with very low value of the baryon load, corresponding to genuine short GRBs. There are also systems as GRB 060614 which do not show a SN \citep{2006Natur.444.1050D,2006Natur.444.1053G} and are expected to be originating possibly from a yet different binary system formed by a white dwarf and a NS \citep{2009A&A...498..501C}. 

Finally, it is appropriate to call on a selection effect in the study of IGC scenario. Only for systems with cosmological redshift $z \lesssim 1$ the current optical instrumentation allows the observation of the related SN Ib/c. A particular challenging analysis is the one of the system GRB 101023 \citep{2012A&A...538A..58P} in which the SN is not detectable but the IGC nature of the source is clearly recognized by the two different episodes in the GRB sources and the spectral features of the first episode.

\begin{acknowledgements}
We thank Massimo Della Valle for multi-years discussions. We are grateful to anonymous referees for important remarks which have improved the presentation of our results.
\end{acknowledgements}

\bibliographystyle{aa}
\bibliography{letter}

\begin{thebibliography}{36}
\expandafter\ifx\csname natexlab\endcsname\relax\def\natexlab#1{#1}\fi

\bibitem[{{Arnett}(1996)}]{1996snai.book.....A}
{Arnett}, D. 1996, {Supernovae and nucleosynthesis. an investigation of the
  history of matter, from the Big Bang to the present}

\bibitem[{{Belvedere} {et~al.}(2012){Belvedere}, {Pugliese}, {Rueda},
  {Ruffini}, \& {Xue}}]{belvedere2012}
{Belvedere}, R., {Pugliese}, D., {Rueda}, J.~A., {Ruffini}, R., \& {Xue}, S.-S.
  2012, Nuclear Physics A, 883, 1

\bibitem[{{Caito} {et~al.}(2009){Caito}, {Bernardini}, {Bianco}, {Dainotti},
  {Guida}, \& {Ruffini}}]{2009A&A...498..501C}
{Caito}, L., {Bernardini}, M.~G., {Bianco}, C.~L., {et~al.} 2009, \aap, 498,
  501

\bibitem[{{Cano} {et~al.}(2011){Cano}, {Bersier}, \& {Guidorzi et
  al.}}]{2011MNRAS.413..669C}
{Cano}, Z., {Bersier}, D., \& {Guidorzi et al.}, C. 2011, \mnras, 413, 669

\bibitem[{{Chevalier}(1989)}]{1989ApJ...346..847C}
{Chevalier}, R.~A. 1989, \apj, 346, 847

\bibitem[{{Chevalier}(2012)}]{2012ApJ...752L...2C}
{Chevalier}, R.~A. 2012, \apj, 752, L2

\bibitem[{{Davidson} \& {Ostriker}(1973)}]{1973ApJ...179..585D}
{Davidson}, K. \& {Ostriker}, J.~P. 1973, \apj, 179, 585

\bibitem[{{Della Valle}(2011)}]{2011IJMPD..20.1745D}
{Della Valle}, M. 2011, International Journal of Modern Physics D, 20, 1745

\bibitem[{{Della Valle} {et~al.}(2006){Della Valle}, {Chincarini}, {Panagia},
  {Tagliaferri}, {Malesani}, {Testa}, {Fugazza}, {Campana}, {Covino},
  {Mangano}, {Antonelli}, {D'Avanzo}, {Hurley}, {Mirabel}, {Pellizza},
  {Piranomonte}, \& {Stella}}]{2006Natur.444.1050D}
{Della Valle}, M., {Chincarini}, G., {Panagia}, N., {et~al.} 2006, \nat, 444,
  1050

\bibitem[{{Gal-Yam} {et~al.}(2006){Gal-Yam}, {Fox}, {Price}, {Ofek}, {Davis},
  {Leonard}, {Soderberg}, {Schmidt}, {Lewis}, {Peterson}, {Kulkarni}, {Berger},
  {Cenko}, {Sari}, {Sharon}, {Frail}, {Moon}, {Brown}, {Cucchiara}, {Harrison},
  {Piran}, {Persson}, {McCarthy}, {Penprase}, {Chevalier}, \&
  {MacFadyen}}]{2006Natur.444.1053G}
{Gal-Yam}, A., {Fox}, D.~B., {Price}, P.~A., {et~al.} 2006, \nat, 444, 1053

\bibitem[{{Gursky} \& {Ruffini}(1975)}]{1975ASSL...48.....G}
{Gursky}, H. \& {Ruffini}, R., eds. 1975, Astrophysics and Space Science
  Library, Vol.~48, {Neutron stars, black holes and binary X-ray sources;
  Proceedings of the Annual Meeting, San Francisco, Calif., February 28, 1974}

\bibitem[{{Hjorth} \& {Bloom}(2011)}]{2011arXiv1104.2274H}
{Hjorth}, J. \& {Bloom}, J.~S. 2011, arXiv:1104.2274

\bibitem[{{Immler} {et~al.}(2002){Immler}, {Wilson}, \&
  {Terashima}}]{2002ApJ...573L..27I}
{Immler}, S., {Wilson}, A.~S., \& {Terashima}, Y. 2002, \apj, 573, L27

\bibitem[{{Iwamoto} {et~al.}(1994){Iwamoto}, {Nomoto}, {Hoflich}, {Yamaoka},
  {Kumagai}, \& {Shigeyama}}]{1994ApJ...437L.115I}
{Iwamoto}, K., {Nomoto}, K., {Hoflich}, P., {et~al.} 1994, \apj, 437, L115

\bibitem[{{Izzo} {et~al.}(2012){Izzo}, {Ruffini}, {Penacchioni}, {Bianco},
  {Caito}, {Chakrabarti}, {Rueda}, {Nandi}, \&
  {Patricelli}}]{2012A&A...543A..10I}
{Izzo}, L., {Ruffini}, R., {Penacchioni}, A.~V., {et~al.} 2012, \aap, 543, A10

\bibitem[{{Leach} \& {Ruffini}(1973)}]{1973ApJ...180L..15L}
{Leach}, R.~W. \& {Ruffini}, R. 1973, \apj, 180, L15

\bibitem[{{Muccino} {et~al.}(2012){Muccino}, {Ruffini}, {Bianco}, {Izzo}, \&
  {Penacchioni}}]{2012arXiv1205.6600M}
{Muccino}, M., {Ruffini}, R., {Bianco}, C.~L., {Izzo}, L., \& {Penacchioni},
  A.~V. 2012, arXiv:1205.6600

\bibitem[{{Negreiros} {et~al.}(2012){Negreiros}, {Ruffini}, {Bianco}, \&
  {Rueda}}]{2012A&A...540A..12N}
{Negreiros}, R., {Ruffini}, R., {Bianco}, C.~L., \& {Rueda}, J.~A. 2012, \aap,
  540, A12

\bibitem[{{Nomoto} \& {Hashimoto}(1988)}]{1988PhR...163...13N}
{Nomoto}, K. \& {Hashimoto}, M. 1988, \physrep, 163, 13

\bibitem[{{Nomoto} {et~al.}(1994){Nomoto}, {Yamaoka}, {Pols}, {van den Heuvel},
  {Iwamoto}, {Kumagaiparallel}, \& {Shigeyama}}]{nomoto1994}
{Nomoto}, K., {Yamaoka}, H., {Pols}, O.~R., {et~al.} 1994, Nature, 371, 227

\bibitem[{{Penacchioni} {et~al.}(2012){Penacchioni}, {Ruffini}, {Izzo},
  {Muccino}, {Bianco}, {Caito}, {Patricelli}, \& {Amati}}]{2012A&A...538A..58P}
{Penacchioni}, A.~V., {Ruffini}, R., {Izzo}, L., {et~al.} 2012, \aap, 538, A58

\bibitem[{{Rawls} {et~al.}(2011){Rawls}, {Orosz}, {McClintock}, {Torres},
  {Bailyn}, \& {Buxton}}]{2011ApJ...730...25R}
{Rawls}, M.~L., {Orosz}, J.~A., {McClintock}, J.~E., {et~al.} 2011, \apj, 730,
  25

\bibitem[{{Rueda} \& {Ruffini}(2012)}]{2012ApJ...758L...7R}
{Rueda}, J.~A. \& {Ruffini}, R. 2012, \apj, 758, L7

\bibitem[{{Ruffini}(2011)}]{2011IJMPD..20.1797R}
{Ruffini}, R. 2011, International Journal of Modern Physics D, 20, 1797

\bibitem[{{Ruffini} {et~al.}(2005){Ruffini}, {Bernardini}, \& {Bianco et
  al.}}]{2005tmgm.meet..369R}
{Ruffini}, R., {Bernardini}, M.~G., \& {Bianco et al.}, C.~L. 2005, in The
  Tenth Marcel Grossmann Meeting. On recent developments in theoretical and
  experimental general relativity, gravitation and relativistic field theories,
  ed. M.~{Novello}, S.~{Perez Bergliaffa}, \& R.~{Ruffini} (Singapore: World
  Scientific), 369

\bibitem[{{Ruffini} {et~al.}(2008){Ruffini}, {Bernardini}, \& {Bianco et
  al.}}]{2008mgm..conf..368R}
{Ruffini}, R., {Bernardini}, M.~G., \& {Bianco et al.}, C.~L. 2008, in The
  Eleventh Marcel Grossmann Meeting, ed. H.~{Kleinert}, R.~T. {Jantzen}, \&
  R.~{Ruffini}, 368--505

\bibitem[{{Ruffini} {et~al.}(2001){Ruffini}, {Bianco}, {Fraschetti}, {Xue}, \&
  {Chardonnet}}]{2001ApJ...555L.117R}
{Ruffini}, R., {Bianco}, C.~L., {Fraschetti}, F., {Xue}, S.-S., \&
  {Chardonnet}, P. 2001, \apj, 555, L117

\bibitem[{{Ruffini} {et~al.}(2000){Ruffini}, {Salmonson}, {Wilson}, \&
  {Xue}}]{2000A&A...359..855R}
{Ruffini}, R., {Salmonson}, J.~D., {Wilson}, J.~R., \& {Xue}, S.-S. 2000, \aap,
  359, 855

\bibitem[{{Schreier} {et~al.}(1972){Schreier}, {Levinson}, {Gursky}, {Kellogg},
  {Tananbaum}, \& {Giacconi}}]{1972ApJ...172L..79S}
{Schreier}, E., {Levinson}, R., {Gursky}, H., {et~al.} 1972, \apj, 172, L79

\bibitem[{{Sedov}(1946)}]{sedov46}
{Sedov}, L.~I. 1946, Compt. Rend. (Doklady) Acad. Sci. URSS, 52, 17

\bibitem[{{Sedov}(1959)}]{1959sdmm.book.....S}
{Sedov}, L.~I. 1959, {Similarity and Dimensional Methods in Mechanics}

\bibitem[{{Soria} {et~al.}(2004){Soria}, {Pian}, \&
  {Mazzali}}]{2004A&A...413..107S}
{Soria}, R., {Pian}, E., \& {Mazzali}, P.~A. 2004, \aap, 413, 107

\bibitem[{{Tananbaum} {et~al.}(1972){Tananbaum}, {Gursky}, {Kellogg},
  {Levinson}, {Schreier}, \& {Giacconi}}]{1972ApJ...174L.143T}
{Tananbaum}, H., {Gursky}, H., {Kellogg}, E.~M., {et~al.} 1972, \apj, 174, L143

\bibitem[{{Tutukov} \& {Fedorova}(2007)}]{2007ARep...51..291T}
{Tutukov}, A.~V. \& {Fedorova}, A.~V. 2007, Astronomy Reports, 51, 291

\bibitem[{{Wilson}(1972)}]{1972ApJ...174L..27W}
{Wilson}, R.~E. 1972, \apj, 174, L27

\bibitem[{{Woosley} \& {Bloom}(2006)}]{2006ARA&A..44..507W}
{Woosley}, S.~E. \& {Bloom}, J.~S. 2006, \araa, 44, 507

\end{thebibliography}

\end{document}